\documentclass{jfm}

\usepackage{epsfig}
\usepackage{subfigure}
\usepackage{graphicx}
\usepackage{pgf}
\usepackage{float}

\newsavebox{\astrutbox}
\sbox{\astrutbox}{\rule[-5pt]{0pt}{20pt}}

\usepackage{amsmath}

\title[Migration of particulate suspensions in a cylindrical Couette]{Fully developed and transient concentration profiles of particulate suspensions sheared in a cylindrical Couette}

\author[M. ~Sarabian, M.~Firouznia, B.~Metzger \& S.~Hormozi ]%
{Mohammad Sarabian$^1$, Mohammadhossein Firouznia$^1$,\\ Bloen Metzger$^2$ \& Sarah Hormozi$^1$ \thanks{Email address for correspondence: hormozi@ohio.edu}
}

\affiliation{
	$^1$Department of  Mechanical Engineering, Ohio University, 251 Stocker Center, Athens, Ohio, USA, 45701.\\
	$^2$Aix-Marseille Universite, CNRS, IUSTI UMR 7343, 13453 Marseille, France.}

\pubyear{xxx}
\volume{xxx}
\pagerange{xxx--yyy}
\date{?? and in revised form ??}
\begin{document}

\maketitle

 \begin{abstract}
We experimentally investigate particle migration in a non-Brownian suspension sheared in a Taylor-Couette configuration and in the limit of vanishing Reynolds number. Highly resolved index-matching techniques are used to measure the local particulate volume fraction. In this wide-gap Taylor-Couette configuration, we find that for a large range of bulk volume fraction, $\phi_b \in [20\% - 50\%]$, the fully developed concentration profiles are well predicted by the Suspension Balance Model of Nott \& Brady (\textit{J. Fluid Mech.}, vol. 275, 1994, pp. 157–199). Moreover, we provide systematic measurements of the migration strain scale and of the migration amplitude which  highlight the limits of the suspension balance model predictions.\\
\textbf{Key words: }suspensions
\end{abstract}
\section{Introduction}

{Shear}-induced migration occurs in suspensions of particles sheared even at low  {Reynolds} number when the flow configuration presents shear-rate {gradient} such as in pipe flows or in large-gap Taylor-Couette flows. In such geometries, particles migrate towards regions of low shear rate producing an inhomogeneous concentration field. This phenomenology can be particularly problematic in the industrial context as it strongly affects the rheological response of suspensions. It can induce clogging or give rise to inhomogeneous preparations.

Two main descriptions have been proposed to model {shear}-induced migration. The earliest comprehensive work dates back to the phenomenological approach of Acrivos and his co-workers. This approach adds diffusive terms to the continuity equation for solid-phase continuum (\cite{Leighton1987shear, Phillips1992constitutive}). These diffusion terms include the effects of particle collisions, {gradient} in the relative viscosity of the suspension, and Brownian motion. Another approach is the Suspension Balance Model (SBM) of \cite{Nott1994pressure}, which is a continuum description accounting for the relative motion between the fluid and the particle phases. Within this latter model, the particle flux is driven by the presence of particle normal stress gradients.  In viscous suspensions, the particle normal stresses scale linearly with the local shear rate. Thus, in  inhomogeneous flows where the shear rate varies with space, particle normal stress gradients spontaneously arise, thereby leading to particle migration. The closures for stresses in suspensions are based on rheological laws such as those of \cite{Morris1999curvilinear} and of \cite{Boyer2011unifying} which involve many constants and constitutive laws that were determined from computational studies and experimental measurements. 

Shear-induced migration has also been investigated computationally via discrete particle {simulation} methods such as:  Stokesian Dynamics {(SD)}, force coupling method {(FCM)} and immersed boundary method {(IBM)}. We refer the reader to the recent comprehensive review of  \cite{Maxey2017}. 

On the experimental front, measurements were carried out in different geometries such as cylindrical Couette cells (\cite{Leighton1987shear}, \cite{Abbott1991experimental}, \cite{graham1991note}, \cite{Phillips1992constitutive}, \cite{Chow1994shear}, {\cite{corbett1995magnetic}}, \cite{Ovarlez2006local}, \cite{Gholami2018}), pipes (\cite{Karnis1966kinetics}, \cite{Hampton1997migration}, \cite{Oh2015pressure}, \cite{Snook2016dynamics}), channels (\cite{Koh1994experimental}, \cite{Lyon1998experimental}) and parallel disks (\cite{Chow1994shear}, \cite{Deshpande2010particle}). In these works, different experimental techniques have been {employed} such as Nuclear Magnetic Resonance (NMR) imaging, laser Doppler velocimetry (LDV),  refractive index matching (RIM), X-ray radiography and Computed Tomography (CT). In cylindrical geometries, SBM is found to provide satisfactory predictions for the fully developed volume fraction and velocity profiles, where the diffusive flux model encounters difficulties (\cite{stickel2005fluid}). 

However, in such geometries, the data characterizing the migration dynamics are sparse (only available for specific concentrations). Conversely, in pipe geometries, the recent experimental work of \cite{Snook2016dynamics} provides substantial information for both the steady-state profiles and the transient evolution of the migration process. They show that the volume fraction at the pipe centreline (i.e. where the shear rate vanishes) decreases with decreasing {bulk} volume fraction while the SBM predicts a centreline volume fraction reaching maximum packing. Moreover, the SBM predicts a significantly faster rate of migration than what is observed experimentally.  

This short summary shows that the description of migration of non-Brownian suspensions remains  incomplete. In the present paper, we investigate shear-induced migration in a wide-gap Taylor-Couette geometry providing systematic results, \textit{i.e.} for a large range of bulk volume fraction, for both the fully developed and the transient evolution of the concentration profiles. This geometry differs from the Poiseuille configuration of \cite{Snook2016dynamics} in that it contains no region flowing at zero shear rate. We find that, in such a case, SBM provides more accurate predictions for both the fully developed and the transient evolution of the concentration profiles. The highly converged and detailed experimental results provided here allow us to highlight the limits of the SBM and discuss the relevance of the different rheological laws on which this model is grounded. These experimental results could represent a benchmark for fully resolved numerical simulations of curvilinear flows. After presenting the experimental procedure in section \ref{sec:Experimental Procedure}, the results are detailed in section \ref{sec:results}. Conclusions are {drawn} in section \ref{sec: Conclusion}.

\section{Experimental set-up and protocols}
\label{sec:Experimental Procedure}
\begin{figure}
	\centering
	\includegraphics[trim={0cm 0cm 0cm 0cm},width=0.7\linewidth]{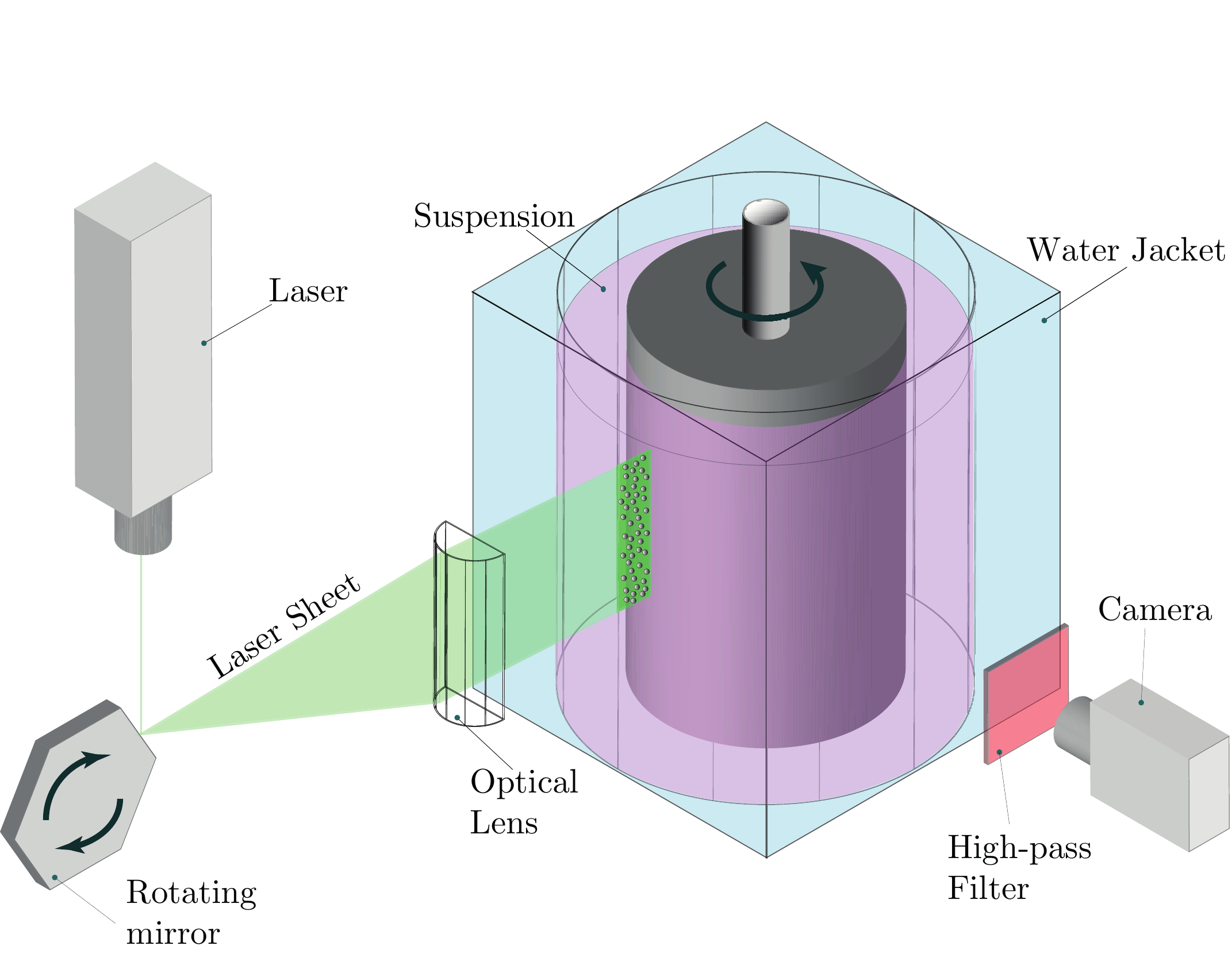}
	\caption{Schematic of the experimental Taylor Couette cell.}
	\label{Experimental-SetUp}
\end{figure}
\subsection{Experimental Apparatus} 
The experimental set-up is sketched in  figure \ref{Experimental-SetUp}. The inner cylinder and outer cylinder of this Couette device have a radius of $R_i=40~mm$ and $R_o=60~mm$ respectively, and their height is 120 mm. The transparent and fixed outer cylinder is machined from a full block of poly(methyl methacrylate) (PMMA) and is embedded in a square water-bath jacket connected to a recirculator (Isotemp 4100R20 Bath-115V), enabling one to control the temperature of the whole set-up ($\pm 0.1^{\circ}C$). The inner moving cylinder is driven by a precision rotating stage (M-061.PD PI piezo-nano positioning) with an angular resolution of $3\times 10^{-5}$ rad. 

We use a semiconductor green laser diode of $200$ mW power and wavelength of $532$ nm as the light source to illuminate a vertical plane ($r,z$) of the suspension. The light beam, after reflection on the rotating mirror, is focused down to a thin ($\approx 60$ $\mu$m) vertical light sheet using a plano-convex cylindrical lens. The suspending fluid is doped with rhodamine 6G which fluoresces under the laser diode illumination. 

\subsection{Particles and fluid}
\label{sec:Particles and fluid and Suspension Preparation}
Particles are rigid PMMA spheres of radius $a=0.79 \pm{0.01} \hspace{1ex}$ mm and density of $1.19 \hspace{1ex} g/cm^3$. The gap to particle size ratio is thus $\ell_{\rm gap}/a=(R_o-R_i)/a\approx25$. 
The suspending fluid is a viscous mixture, similar to that used by {{\cite{Pham2016origin}}}, composed of Triton X-100 ($76.242 \hspace{1ex} wt\%$), Zinc Chloride ($14.561\hspace{1ex} wt\%$), Water ($9.112 \hspace{1ex} wt\%$) and Hydrochloric Acid (HCl) ($0.085 \hspace{1ex} wt\%$), having at room temperature a viscosity of $\eta= 4.64$ Pa s. This large viscosity of the suspending fluid ensures that experiments are performed in the limit of vanishing Reynolds number ($Re<10^{-3}$). A small amount of hydrochloric acid (HCl) is added to the preparation to improve its transparency. The composition of the suspending fluid is chosen to match both the refractive index and the density of the PMMA particles. This makes the suspension transparent and prevents sedimentation of the particles. 

After preparation, the bubble-free suspension is delicately poured into the Couette cell. Further mixing inside the cell is performed using a thin rod to make sure that the initial particle distribution is homogeneous throughout the gap. 
\subsection{Flow visualization}
\label{sec:Refractive Index Matching Procedure}
In order to distinguish particles from the fluid phase while imaging, rhodamine 6G is added to the suspending fluid with a concentration of $ 0.1\hspace{1ex} mg/L$. This amount of dye is chosen to maximize the contrast between the bright fluorescent fluid and the dark particles, while allowing the laser light to penetrate deep enough into the measurement volume.
The temperature of the set-up is set to $20^{\circ}C \pm 0.1^{\circ}C$ in order to optimize index matching between the suspending fluid and the particles (within $\pm 10^{-4}$). This temperature adjustment improves index matching (down to the fourth digit) and thereby the imaging  quality. Image acquisition is performed by using a digital camera (Basler Ace acA2000-165um USB3 Monochromatic) and a high-quality magnification lens (Sigma APO-Macro-180 mm-F3.5-DG). A $550$ nm long-pass coloured glass filter is placed before the lens to remove all direct light reflections and only keep the light re-emitted by the fluorescent fluid. 
\subsection{Experimental protocol and data analysis}
\label{sec:Data Analysis}

\begin{figure}
	\centering
	\includegraphics[trim={0cm 0cm 0cm 0cm},width=0.6\linewidth]{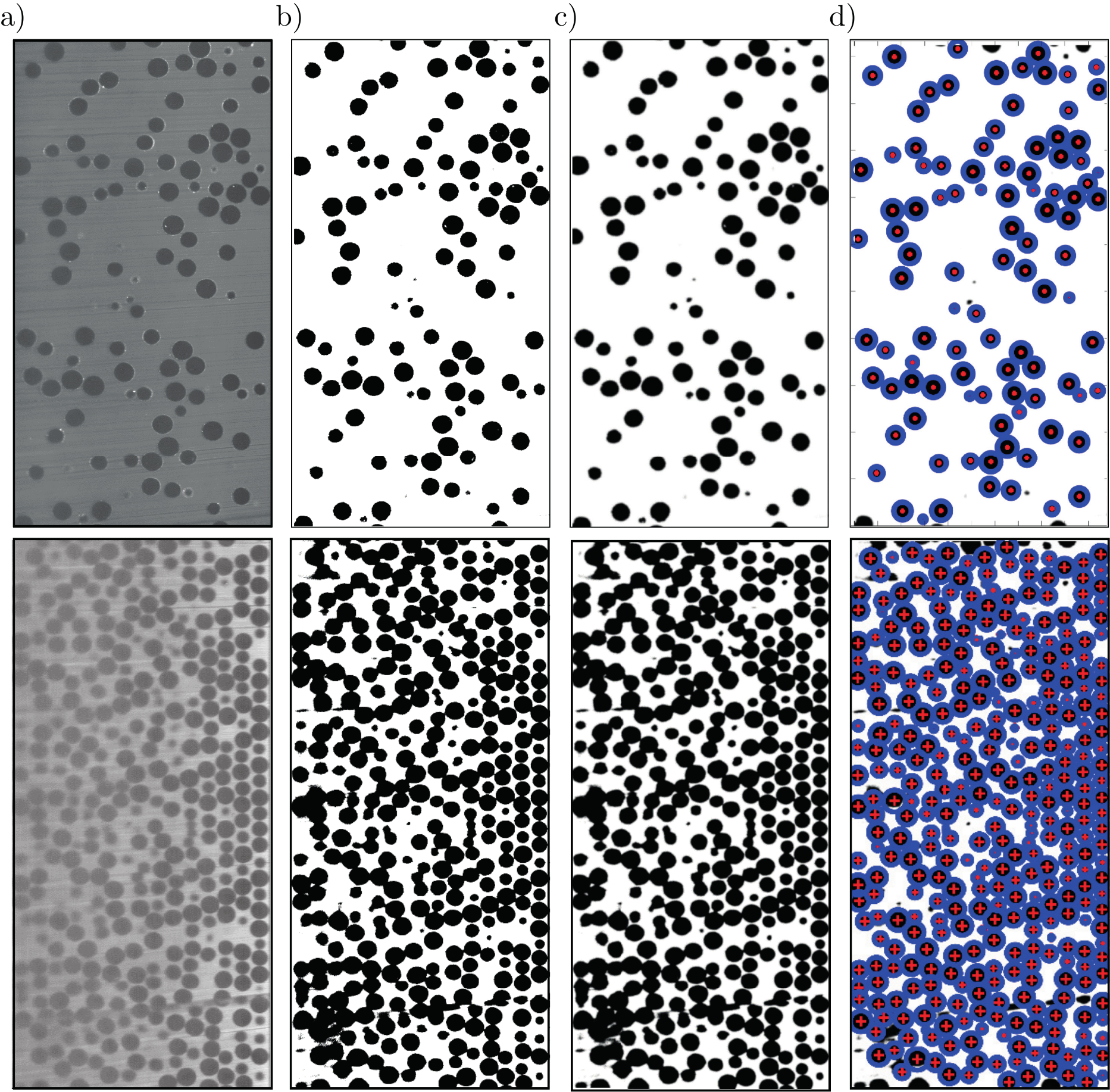}
	\caption{Image processing steps for (top) $\phi_{b}=20\%$ and (bottom) $\phi_{b}=50\%$: a) Raw image, b) Thresholding, c) Gaussian blur filtering, d) Hough transform.}
	\label{ImageAnalysis}
\end{figure}
\begin{figure}
\centering
\includegraphics[trim={0cm 0cm 0cm 0cm},width=1\linewidth]{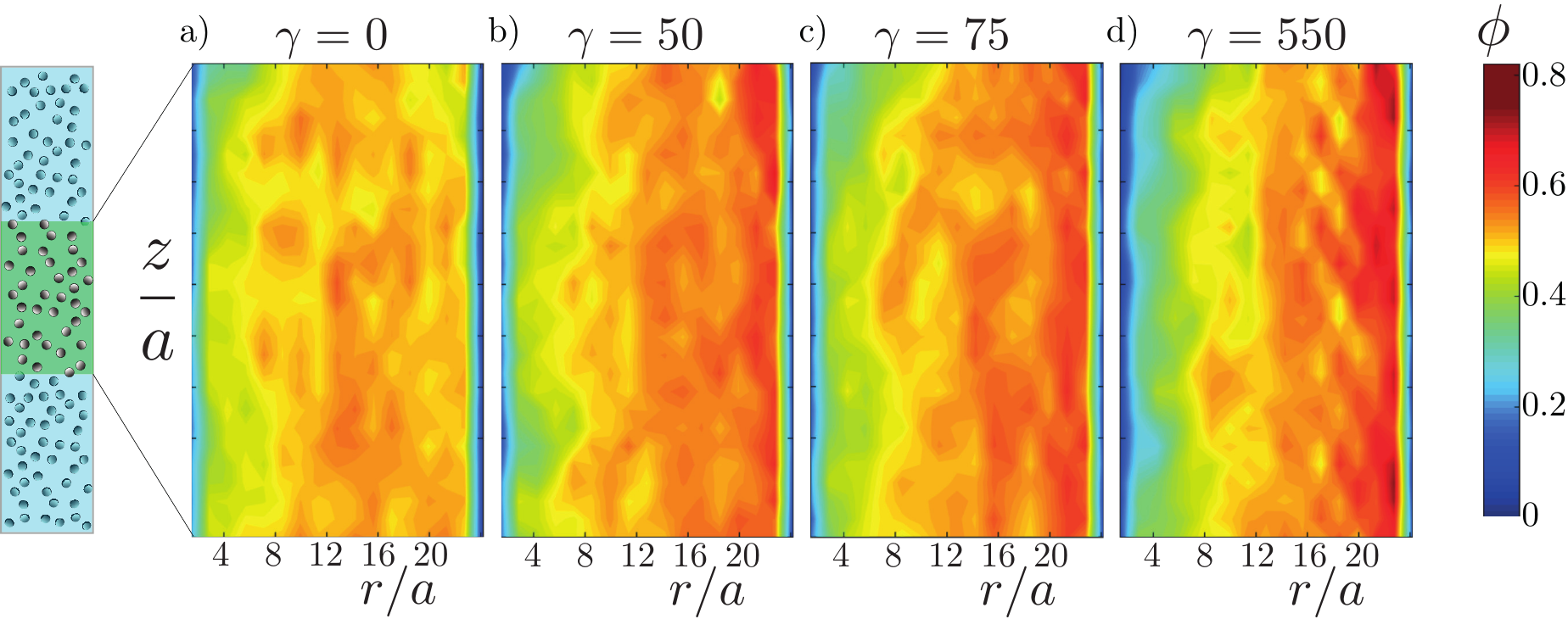} 
\caption{Evolution of the concentration field $\phi(r,z,t)$ for $\phi_{b}=50\%$ at a) $\gamma=0$, b) $\gamma=50$, c) $\gamma=75$ and d) $\gamma=550$ (steady state).}
\label{field}
\end{figure}

Experiments are performed for six different bulk volume fractions, (i.e $20\% \leq \phi_b \leq 50\%$). For each experiment, once the suspension is poured and homogenized into the cell, the suspension is sheared and image acquisition is started. In all of these experiments the rotating speed of the inner cylinder is $\Omega_i=0.058 ~rad/sec$ resulting in a shear rate at the inner cylinder of $\dot\gamma=0.2$ s$^{-1}$.  Images are acquired every strain unit for a total accumulated strain of $4500$ for the lowest volume fraction. To provide a measure of accuracy and error bars, four experiments were performed for each bulk volume fraction following exactly the same procedure.

To obtain the particle concentration field $\phi(r,z,t)$, where $r$ and $z$ correspond to the radial and vorticity direction respectively, we use an image processing method similar to that of Snook et al. (2016). As shown in figure \ref{ImageAnalysis}, the raw images are successively thresholded, blurred and finally, a circular Hough transform is applied to detect each particle centre and apparent radius in the plane of visualization (\cite{Duda1972use}). It is noteworthy to mention that in figure  \ref{ImageAnalysis}, all particles have the same size; their apparently different radii arise from their intersection with the light sheet. Each image is divided into 125x235 overlapping horizontal and vertical sub-images of width $0.1d_p$. The concentration field is obtained by integrating the area covered by the particles within each sub-image. This information can easily be inferred from the knowledge of the particle centre locations and their apparent radii within the visualization sheet. The above technique was found to be more accurate than directly measuring the local volume fraction from the thresholded image, the latter method being biased by unavoidable horizontal streaks on the images arising from dust or air bubbles. 

Since the concentration field evolves slowly, concentration fields obtained for successive images are time averaged over 10 successive images (10 strain units) to reduce fluctuations. Figure \ref{field} shows a typical example of the concentration field measured as a suspension of bulk volume fraction $\phi_b=50\%$ is sheared.  The suspension is initially rather homogeneous and, as it is sheared, the volume fraction decreases near the inner rotating cylinder (small $r$) while it increases near the outer fixed cylinder (large $r$). Note that the last concentration field, \textit{i.e} for figure \ref{field} .d, corresponds to an accumulated strain of $550$ at which the concentration field reached steady state. We find that the concentration averaged over the whole domain $\phi_{\rm average}$ (\textit{i.e.} averaged over $r$ and $z$) is constant in time remaining equal to $\phi_b$, the initial bulk volume fraction; see Figure \ref{Evolution} (Green square). This result indicates that migration in the $z$-direction is negligible, as can also been observed from the invariance along $z$ of the concentration field in Figure \ref{field}.d. We thus spatially average the concentration fields $\phi(r,z,t)$ along $z$ to obtain the concentration profile $\phi(r,t)$.  In the following, these radial concentration profiles are used to characterize the migration process. 
\section{Results}
\label{sec:results}

\begin{figure}
	\centering
	\includegraphics[trim={1.6cm 0cm 0cm 0cm},width=0.9\linewidth]{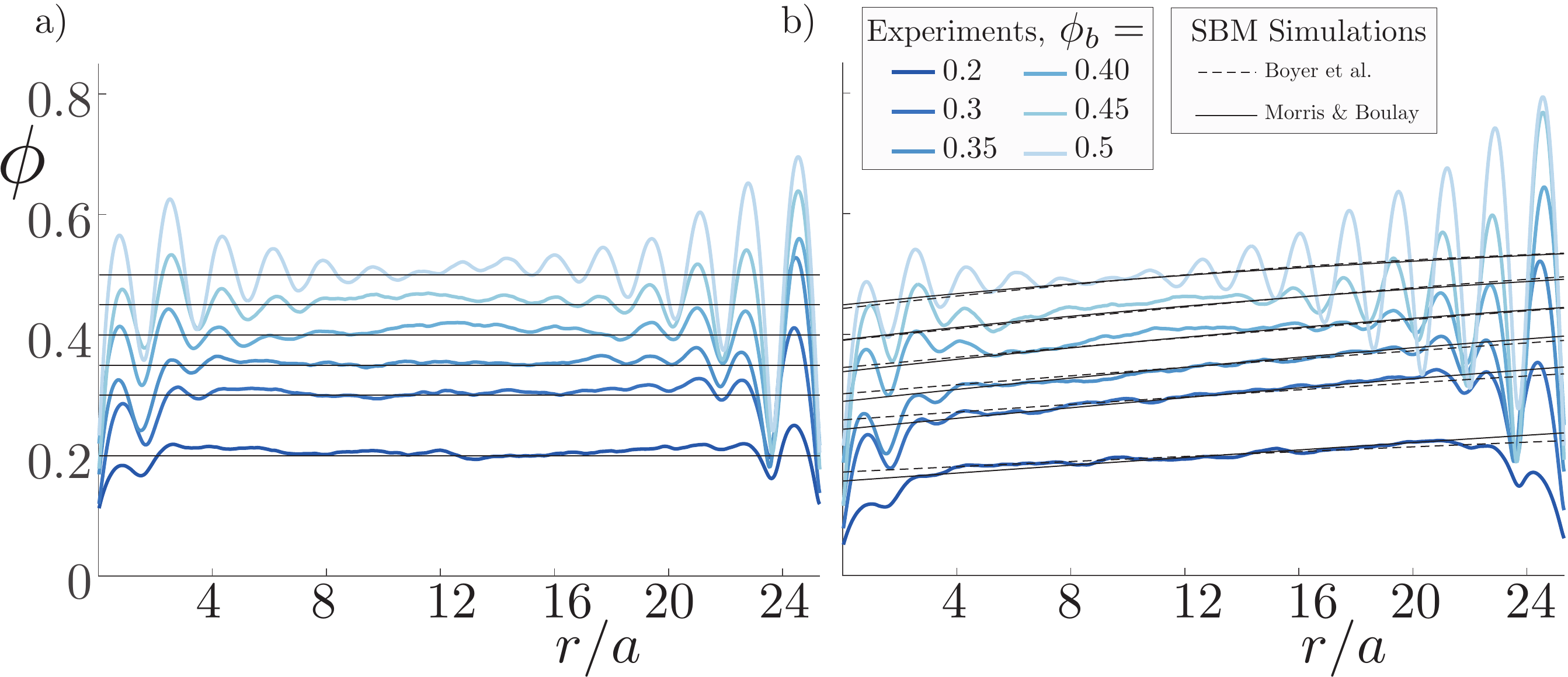}
	\caption{a) Initial and b) steady state concentration profiles averaged over four experimental runs. Black solid and dashed lines correspond to SBM predictions using Morris $\&$ Boulay and Boyer $\&$ co-workers rheology respectively.}
	\label{Initial-Steady} 
\end{figure}

\subsection{Initial and steady-state profiles}
Figure \ref{Initial-Steady} shows the initial and steady-state (fully developed) concentration profiles (averaged over four experimental runs) across the gap of the cell for different bulk volume fractions. The initial particle concentration profiles are homogeneous  across the gap for all the bulk volume fractions.  Some particle layering can be observed in the vicinity of both the inner and outer cylinders especially as the bulk volume fraction is increased. For all the bulk volume fractions investigated, the fully developed concentration profiles clearly indicate a migration of the particles from the inner cylinder (high-shear-rate-region) {towards} the outer cylinder (low-shear-rate-region), see figure \ref{Initial-Steady}.b. Moreover, particle layering is more pronounced as the bulk solid volume fraction increases with a stronger layering close to the outer cylinder.

In figure \ref{Initial-Steady}.b, we compare our experimental results to the predictions of the SBM obtained using the rheological laws of \cite{Morris1999curvilinear} and \cite{Boyer2011unifying}. The SBM framework, the rheological laws and the numerical scheme implemented  are  presented in Appendix \ref{SBM Equations}. We find that the SBM predictions capture well the average migration process and provide very similar results when using \cite{Morris1999curvilinear} or \cite{Boyer2011unifying} laws. Obviously, this model, which is based on a continuum approach, does not capture the observed particle layering which is a discrete effect induced by the finite particle size and confinement (e.g. \cite{alghalibi2018interface}).

\subsection{Migration dynamics}

\begin{figure}
	\centering
		\includegraphics[trim={0cm 0cm 0cm 0cm},width=1\linewidth]{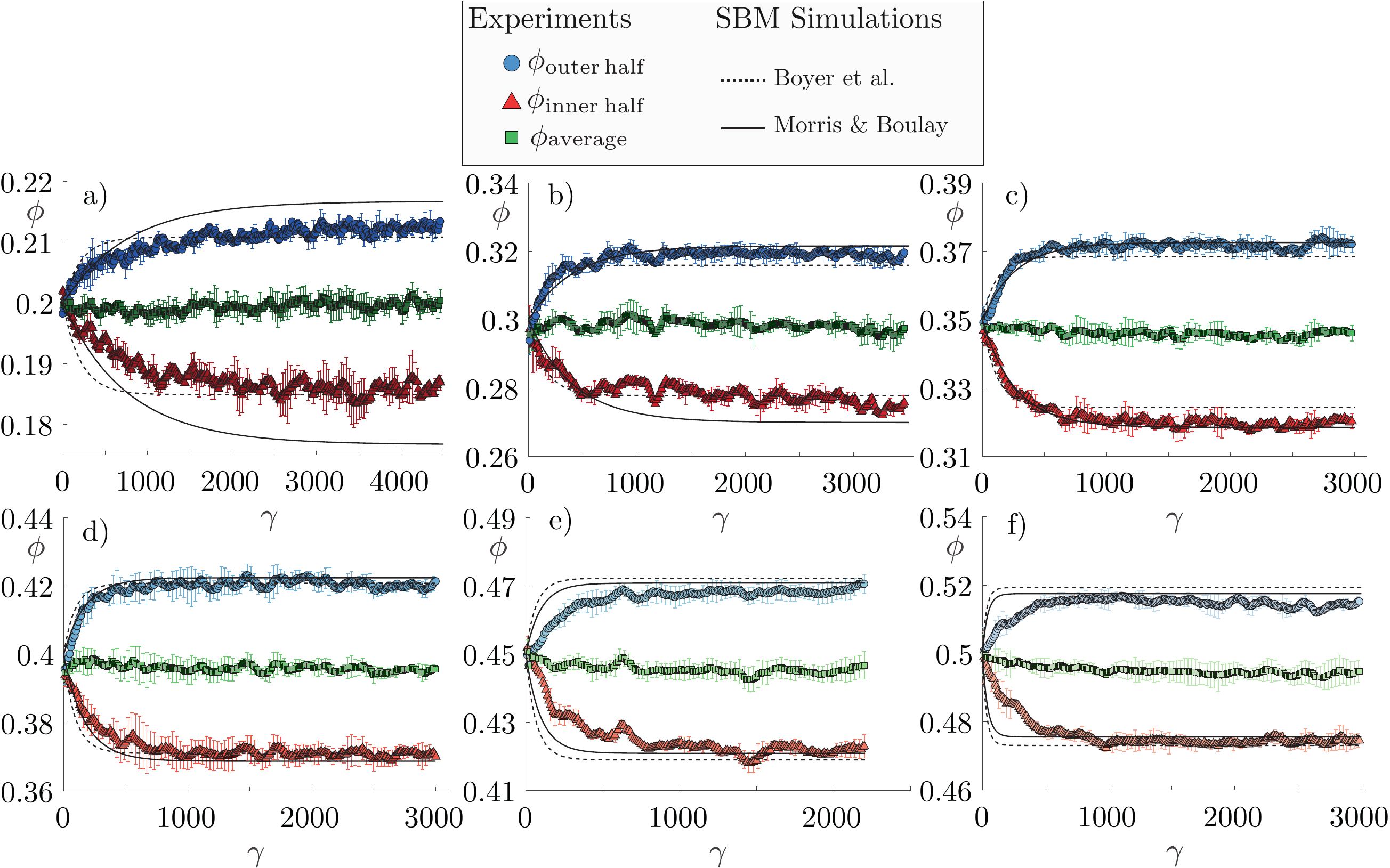}
	\caption{Evolution of the concentration, averaged over the inner half (red triangle),  the outer half (blue circle) and the whole gap (green square) for different initial {bulk} volume fractions: a) $\phi_b=20\%$ b) $\phi_b=30\%$, c) $\phi_b=35\%$ d) $\phi_b=40\%$, e) $\phi_b=45\%$ f) $\phi_b=50\%$. Comparison with SBM using (full line) Morris $\&$ Boulay and (dashed line) Boyer rheology.}
	\label{Evolution}
\end{figure}
To investigate the dynamics of the migration process, we compute the average value of the concentration over the inner half of the gap $\phi_{\rm \scriptsize inner-half}$ (from the inner cylinder to the middle of the gap) and over the outer half $\phi_{\rm \scriptsize outer-half}$ (from the middle of the gap to the outer cylinder). Figure \ref{Evolution} shows the evolution of these quantities averaged over four experimental runs as a function of the accumulated strain $\gamma$ for experiments performed with different bulk volume fractions.  Starting from $\phi_b$, the volume fraction averaged over the inner half of the gap (where the shear rate is the largest) decreases while that of the outer half (where the shear rate is the lowest) increases before eventually reaching a steady-state value. Again we compare these experimental measurements to the predictions obtained from the SBM solved using both rheological laws of  \cite{Morris1999curvilinear} and \cite{Boyer2011unifying}.  A striking result is that, conversely to what was found in the experiments of \cite{Snook2016dynamics} performed in a Poiseuille configuration, here in a Taylor-Couette configuration, the SBM predictions of the transient evolution seem to show a better agreement with the experimental results. The inspection of these curves indicates that the SBM migration strain scale is in rather good agreement at low volume fraction. However the SBM predicts a faster migration at large volume fraction than that observed experimentally.

\begin{figure}
\centering
\includegraphics[trim={0cm 0cm 0cm 0cm},width=0.8\linewidth]{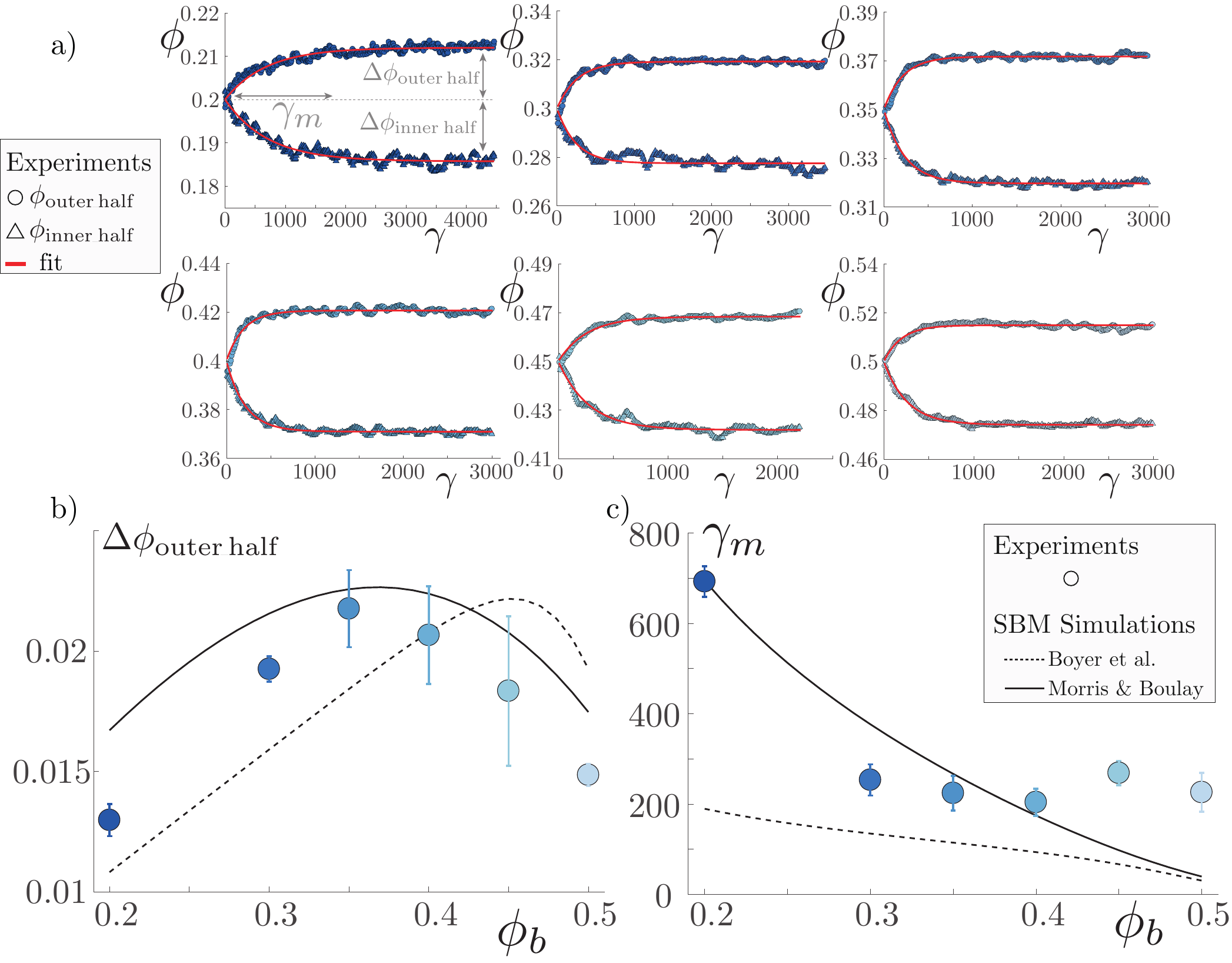} 	
\caption{a) Inner and outer mean volume fractions versus strain fitted (red lines) with $\phi_b-\Delta \phi_{\rm \scriptsize inner-half}(1-e^{-\gamma/\gamma_m})$ and $\phi_b+\Delta \phi_{\rm \scriptsize outer-half} (1-e^{-\gamma/\gamma_m})$ respectively, for different initial bulk volume fractions $\phi_b=20, 30, 35, 40, 45, 50\%$. b) Amplitude of migration $\Delta \phi_{\rm \scriptsize outer-half}$ versus initial bulk volume fraction $\phi_b$ (The same trends are observed with $\Delta \phi_{\rm \scriptsize inner-half}$ ). c) Migration strain scale $\gamma_m$ to reach steady state versus bulk volume fraction $\phi_b$. Comparison with SBM using (full line) Morris $\&$ Boulay and (dashed line) Boyer rheology.}
\label{EvolutionFitting}
 \end{figure}

To provide a more quantitative comparison between the SBM predictions and our experimental results, we investigate the migration dynamics by fitting exponential relaxations of the form $\phi=\phi_b\pm\Delta \phi (1-e^{-\gamma/\gamma_m})$ to the transient evolution of the volume fraction in the outer-half part of the gap; see Figure \ref{EvolutionFitting}.a. Here, as depicted in Figure \ref{EvolutionFitting}.a, $\Delta \phi $ corresponds to the variation of the volume fraction within the outer-half part of the gap and $\gamma_m$ is the typical strain scale for migration to proceed.

Interestingly, as shown in Figure \ref{EvolutionFitting}.b, we find that the strength of the migration process, here characterized by $\Delta \phi_{\rm \scriptsize outer-half}$, is a non-monotonous function of $\phi_b$. Note that the same trend is found for $\Delta \phi_{\rm \scriptsize inner-half}$ (yet not plotted for the sake of clarity). This non-monotonic behaviour can easily be understood since, in both limits where $\phi_b \rightarrow 0$ and $\phi_b \rightarrow \phi_c$ ($\phi_c$ corresponding to the critical volume fraction at which the suspension jams) no migration should occur, \textit{i.e.} $\Delta \phi \rightarrow 0$. We thus expect a maximum $\Delta \phi$ in between. In our geometry, we find migration is the strongest for $\phi_b\approx 35\%$.

The above measurement of the amplitude of migration $\Delta \phi$ is used to quantitatively test the SBM predictions. As shown in Figure \ref{EvolutionFitting}.b, both the rheological laws of \cite{Morris1999curvilinear} and  of \cite{Boyer2011unifying} predict a non-monotonic behaviour of $\Delta \phi$. However, that of \cite{Morris1999curvilinear} provides better agreement with the experimental results for the amplitude of migration. It also captures more accurately the bulk volume fraction at which migration is the strongest, here $\phi_b\approx 35\%$.

The fitting by an exponential relaxation also gives us a measure of the migration strain scale $\gamma_m$. Figure \ref{EvolutionFitting}.c shows that $\gamma_m$ decreases when the bulk volume fraction is increased. Again, we find that the rheological law of \cite{Morris1999curvilinear} provides better agreement especially at low volume fractions (when $\phi_b<40\%$), whereas the migration strain scale obtained with \cite{Boyer2011unifying} is too small by approximately a factor 2 to 3. Interestingly, when $\phi_b>30\%$, the critical strain measured experimentally seems to saturate to a value of $\gamma_m\approx 200$, while the SBM predictions keep decreasing. As discussed in the following, this behaviour may be the results of the strong particle layering observed at large volume fractions in our experiments. 

 \section{Conclusion}
 \label{sec: Conclusion}
 
The fully developed and transient particulate volume fraction profiles of non-Brownian suspensions were investigated experimentally in a Taylor-Couette device using refractive index matching and fluorescence imaging techniques. Systematic experiments were performed for a large range of bulk volume fractions ranging from $20\%$ to $50\%$. Steady-state concentration profiles as well as their dynamics were analysed and compared with the predictions of the suspension balance model (SBM) using two different rheological laws introduced by \cite{Morris1999curvilinear} and \cite{Boyer2011unifying}. The present experimental data may be used as a benchmark for refining the continuum SBM framework or discrete-particle simulations.

The comparison of the steady-state profiles obtained experimentally and numerically using the SBM framework provides rather good agreement regardless of the chosen rheological law \cite{Morris1999curvilinear} or \cite{Boyer2011unifying}. However, by performing more systematic characterization of the migration dynamics, we could highlight significant differences, in particular through the measurement of the strain scale $\gamma_m$ needed to reach steady state and of the migration amplitude $\Delta \phi$. We showed that the rheological law of \cite{Morris1999curvilinear} captures more accurately the amplitude and the non-monotonic trend of the migration amplitude $\Delta \phi$. In our geometry, migration is found to be the strongest for $\phi_b\approx 35\%$. Moreover, for moderate volume fractions ($\phi_b<40\%$), the latter rheological law provides quantitative predictions for the migration strain scale, $\gamma_m$, while that predicted by \cite{Boyer2011unifying} is too small by approximately a factor 3. The two rheological laws investigated here mostly differ in the contact contribution to the particle stress term (see appendix A). In Boyer's rheology, this term is approximately $7$ times larger (mostly because of the small value of $I_0$), thereby yielding a faster migration dynamics. It is important to recall that the rheology of \cite{Boyer2011unifying} was obtained experimentally for volume fractions $\phi_b > 40\%$ and that of \cite{Morris1999curvilinear} was developed to describe migration also at large volume fraction. Thus, extending these laws to low volume fraction might require some modifications. However, since our aim here is not to provide new rheological laws but to contribute to the understanding of shear-induced migration, we have plotted the SBM predictions based on these laws using their original form, \textit{i.e.}  without adjusting any parameter. Following this approach, our result shows that the rheological law of \cite{Morris1999curvilinear} captures more accurately the migration process.
 
The highly converged and detailed experimental results provided here should help further to improve the continuum modelling for migration. Indeed, it appears that the SBM framework is globally more successful when applied to cylindrical Couette configurations than to Poiseuille configurations. For instance, the SBM predicts complete jamming at the centreline of a pipe ($\phi \rightarrow \phi_c$) while this trend is usually not observed in experiments (\cite{Snook2016dynamics}). In the same geometry, the predicted migration strain scales are significantly smaller (more than an order of magnitude) than those measured experimentally (\cite{Snook2016dynamics}). Conversely, in the present cylindrical Couette configuration, the steady-state volume fraction profiles and the migration strain scale are found to be captured more accurately. 
One reason that may explain this difference is that the above two systems differ in the fact that in Poiseuille configurations  the shear rate goes to zero at the centreline. In such a region, as a result of migration, the suspension viscosity is expected to diverge as $\phi \rightarrow \phi_c $. The absence of such a singular region in the cylindrical Couette configuration may explain the relative success of SBM in predicting the migration process.

It is also important to discuss two related effects, namely those of \emph{confinement} and \emph{layering}. The present experiments were performed for $\ell_{\rm gap}/a \approx 25$. Such confinement is motivated by several points. In practice, index matching and direct optical visualization require small systems. Moreover, providing experimental results for such confinement is useful to compare with numerical simulations which, limited by computation time, also investigate systems having similar sizes (\cite{Gallier2016}). Most importantly, for $\ell_{\rm gap}/a > 20$, previous experiments and numerical simulations both already showed that confinement has a negligible impact on macroscopic quantities such as the suspension viscosity (\cite{Maxey2017}), but also on the fully developed concentration profiles (\cite{Snook2016dynamics}). We thus performed experiments for  only one confinement $\ell_{\rm gap}/a \approx 25$ for which no systematic bias is expected. Yet, it would be interesting in future studies to vary this parameter and investigate its impact on the migration strain scale.  

The impact of layering is less clear. Layering is inherent to monodisperse suspensions sheared at large volume fraction and it is found to be more prominent in highly confined systems. When increasing $\ell_{\rm gap}/a$ (which damps the layering), \cite{Snook2016dynamics} found similar fully developed concentration profiles.  We also find quite an accurate match between our steady-state concentration profiles and the SBM prediction, even at large volume fraction where layering becomes very strong.  Thus, layering {seems} not to influence the fully developed concentration profiles. However, the impact of layering on the migration dynamics still needs to be clarified. The work of \cite{Metzger2013} shows that at large volume fraction, the structuration of the suspension into layers strongly damps the particle shear-induced diffusion coefficient. As particles organize into layers, fewer collisions per unit of strain occur between particles thereby reducing the amount of fluctuations. According to this scenario, layering should slow down the migration process, as we observed in the present study where, above $\phi_b=40\%$, the migration strain scale no longer decreases. Conversely, the numerical simulations of \cite{Gallier2016} indicate that the particle stress (whose divergence in inhomogeneous flow drives the migration) is only slightly affected by the structuration of the suspension. From the latter observation, one would expect the migration dynamics to be unaffected by layering.  {Whether} layering affects the migration dynamics thus remains an open question. Addressing this point would deserve a dedicated investigation as disrupting the structuration of the suspension requires using polydisperse suspensions. In such a case, one should account for changes in the critical volume fraction of the suspension.

\vspace{-0.5cm}\section*{Acknowledgments}
This work was supported by NSF (Grant No. CBET-1554044-CAREER) and ACS PRF (Grant No. 55661-DNI9) via the research awards (S.H.) and by ANR JCJC SIMI 9, the Labex MEC ANR-10-LABX-0092 and ANR-11-IDEX-0001-02.

\vspace{-0.5cm}
\appendix
\section{Suspension Balance Model (SBM)}\label{App: A}
\label{SBM Equations}
\subsection{Mathematical Modeling} 

In the SBM of  \cite{Nott1994pressure}, particle diffusive fluxes arise from gradients in the particle normal stress, which enter the solid-phase continuity equation directly by considering the particle drag closure.  The transport equation for the solid volume fraction is 
\begin{eqnarray}
\frac{\partial \phi }{\partial {t}} + {\mathbf{\nabla}} \cdot [ \phi  {\mathbf{u}} ]
&=&
- {\mathbf{\nabla}} \cdot [ \phi (1-\phi){M} ~
{\mathbf{\nabla}} \cdot {\Sigma}^p  ], ~M=\frac{d_p^2}{18\eta_{f}\phi(1-\phi)}f(\phi). \label{eq:solids_mass3}
\end{eqnarray}
Where $u$ is the volume average velocity of the suspension and ${\Sigma}^p$ is the particle phase stress. Here, $M$ denotes the particle mobility and $f(\phi)=(1-\phi)^{\alpha}$ is the sedimentation hindrance function (\cite{Richardson1954sedimentation}).  The viscosity of the suspending fluid is denoted by $\eta_{f}$.  The above transport equation for $\phi$ is solved coupled with the equations of continuity and momentum for the suspension given respectively  as follows:
\begin{equation} \label{continuity}
\nabla. \textbf{u}=0, ~~~~\nabla. \Sigma=0,
\end{equation} 
where $ \Sigma$ is the bulk average suspension stress, which is composed of the  particle-phase ($\Sigma^{p}$) and fluid-phase stresses ($\Sigma^{f}$):
\begin{equation} \label{fluidstress}
\Sigma^{f}=-p_{f} \textbf{I}+2\eta_{f}\textbf{E},~~~~ \Sigma^{p}=-\Sigma^{p}_{nn}+2\eta_{f}\eta_{p}\textbf{E}.
\end{equation}
Here,  $p_{f}$ is the fluid-phase pressure, $\textbf{I}$ is the identity tensor, $\textbf{E}$ is the rate-of-strain tensor, and $\Sigma^{p}_{nn}$ is the diagonal component of $\Sigma^{p}$ representing the particle-phase normal stress given by 
\begin{equation} \label{particlenormalstress}
\Sigma^{p}_{nn}=\eta_{f}\eta_{N}|\dot{\gamma}|\textbf{Q}, ~~~~~~\textbf{Q} = \begin{pmatrix} \lambda_{2} & 0 & 0\\ 0 & 1 & 0\\ 0 & 0 & \lambda_{3} \end{pmatrix},
\end{equation}
%
with $\lambda_{2}=\frac{\Sigma^{p}_{22}}{\Sigma^{p}_{11}}$ and $\lambda_{3}=\frac{\Sigma^{p}_{33}}{\Sigma^{p}_{11}}$ (\cite{Dbouk2013shear}). Combining equations  (\ref{fluidstress}) and (\ref{particlenormalstress}), the bulk average suspension stress can be written as  $\Sigma=\Sigma^f+\Sigma^p=-p_{f} \textbf{I}-\eta_{f}\eta_{N}|\dot{\gamma}|\textbf{Q}+ 2\eta_{f}\eta_s\textbf{E} $, Where $\eta_s=\eta_p+1$ and $\eta_N$ are the relative shear and normal viscosity of the suspension respectively. We use the following rheological laws introduced by \cite{Morris1999curvilinear} and  \cite{Boyer2011unifying} to  close the system of equations (\ref{eq:solids_mass3} and \ref{continuity}). We then solve the equations following the numerical scheme of the two-stage, two-level finite difference method proposed by \cite{Meek1982two}.\newline
{\bf \cite{Morris1999curvilinear} rheological law: }
$\eta_{N}=K_{n}(\frac{\phi}{\phi_{m}})^2(1-\frac{\phi}{\phi_{m}})^{-2}$, $\eta_{s}=\frac{\eta}{\eta_{f}}=1+2.5\phi(1-\frac{\phi}{\phi_{m}})^{-1}+K_{s}(\frac{\phi}{\phi_{m}})^2(1-\frac{\phi}{\phi_{m}})^{-2}$, $K_n=0.75$, $K_s=0.1$, $\alpha=4$, $\lambda_{2}=0.8$, $\lambda_{3}=0.5$ and $\phi_{m}=0.585$. \newline
{\bf \cite{Boyer2011unifying}  rheological law:} 
$\eta_{N}=(\frac{\phi}{\phi_{m}})^2(1-\frac{\phi}{\phi_{m}})^{-2}$, $\eta_{s}=\frac{\eta}{\eta_{f}}=1+2.5\phi(1-\frac{\phi}{\phi_{m}})^{-1}+\mu^{c}(\phi)(\frac{\phi}{\phi_{m}})^2(1-\frac{\phi}{\phi_{m}})^{-2}$, 
$\mu^{c}(\phi)=\mu_{1}+\frac{\mu_{2}-\mu_{1}}{[1+I_{0}\phi^{2}(\phi_{m}-\phi)^{-2}]}$,  $\mu_{1}=0.32$, $\mu_{2}=0.7$, $I_{0}=0.005$, $\phi_{m}=0.585$, $\alpha=5$, $\lambda_{2}=0.95$ and $\lambda_{3}=0.6$.
{\subsection{Numerical Algorithm}\label{App: A-Numerical Algorithm}
We solve the system of equations (\ref{eq:solids_mass3}) and (\ref{continuity}) coupled with the rheological laws of \cite{Morris1999curvilinear} and   \cite{Boyer2011unifying} following the numerical scheme of the two-stage, two-level finite difference method proposed by \cite{Meek1982two} . We have used the simplified Newton modified predict-and-correct scheme (MPC). It was shown that this scheme is consistent and second-order-accurate in space. This scheme basically linearizes the standard Crank-Nicolson discretization equation and solves the system of linear algebraic equations at each time step. For more information about the details of the MPC scheme, we refer the reader to \cite{Meek1982two}.}

\end{document}